\documentclass[runningheads,english]{llncs}

\usepackage[utf8]{inputenc}
\usepackage[T1]{fontenc}
\usepackage{microtype}
\usepackage{babel}
\usepackage{csquotes}
\usepackage{tabularx}
\usepackage{booktabs}
\usepackage{varwidth}

\usepackage{amsmath, amssymb} % amsthm removed as llncs defines its own theorem environments
\usepackage{graphicx}
\usepackage{float}
\usepackage{xcolor}
\usepackage{colortbl}
\usepackage{listings}
\usepackage[colorlinks=true, allcolors=blue]{hyperref}
\usepackage[backend=biber, style=lncs, sorting=none]{biblatex}
\usepackage[most]{tcolorbox}
\usepackage{algorithm}
\usepackage{algpseudocode}
\graphicspath{{./}{figures/}} % Include figures/ for arXiv submission

\tcbset{
  verquerdbox/.style={
    enhanced,
    breakable,
    attach boxed title to top left={xshift=3mm, yshift=-3mm, yshifttext=-1mm},
    boxed title style={
      colback=#1,
      colframe=#1,
      arc=0mm,
    },
    fonttitle=\bfseries\small,
    arc=0mm,
    outer arc=0mm,
    boxrule=0.5pt,
    top=4mm,
    bottom=2mm,
    left=2mm,
    right=2mm,
  }
}

\newtcolorbox{infobox}[1]{
  verquerdbox=blue!75!black,
  colback=blue!5!white,
  colframe=blue!75!black,
  title=#1,
}

\newtcolorbox{insight}[1]{
  verquerdbox=green!60!black,
  colback=green!5!white,
  colframe=green!60!black,
  title=#1,
}

\newtcolorbox{examplebox}[1]{
  verquerdbox=orange!80!black,
  colback=orange!5!white,
  colframe=orange!80!black,
  title=#1,
}

\newtcolorbox{pedagogy}[1]{
  verquerdbox=cyan!70!black,
  colback=cyan!5!white,
  colframe=cyan!70!black,
  title=#1,
}

\definecolor{primecolor}{RGB}{200,230,255}
\definecolor{codegreen}{rgb}{0,0.6,0}
\definecolor{codegray}{rgb}{0.5,0.5,0.5}
\definecolor{codepurple}{rgb}{0.58,0,0.82}
\definecolor{backcolour}{rgb}{0.95,0.95,0.92}
\definecolor{deepblue}{rgb}{0,0,0.6}
\definecolor{deepred}{rgb}{0.6,0,0}

\lstdefinestyle{verquerdstyle}{
    backgroundcolor=\color{backcolour},   
    commentstyle=\color{codegreen},
    keywordstyle=\color{blue}\bfseries,
    numberstyle=\tiny\color{codegray},
    stringstyle=\color{deepred},
    basicstyle=\ttfamily\small,
    breakatwhitespace=false,         
    breaklines=true,                 
    captionpos=b,                    
    keepspaces=true,                 
    numbers=left,                    
    numbersep=10pt,                  
    showspaces=false,                
    showstringspaces=false,
    showtabs=false,                  
    tabsize=2,
    frame=none,
    xleftmargin=15pt,
    literate=
      {Ä}{{\"A}}1
      {Ö}{{\"O}}1
      {Ü}{{\"U}}1
      {ä}{{\"a}}1
      {ö}{{\"o}}1
      {ü}{{\"u}}1
      {ß}{{\ss}}1
}
\begin{document}

\title{Adversarial Robustness in Fake Quantum Simulators}

\author{Marc Mau\ss{}ner\inst{1} \and Volker Reers\inst{2}}
\authorrunning{M. Mau\ss{}ner and V. Reers}

\institute{infoteam Software AG \and Qseidon GmbH}

\maketitle

\begin{abstract}
This paper investigates the performance scalability and adversarial robustness of Quantum Machine Learning (QML) models deployed on noise-model-based fake simulators. We conduct a dual-phased study, first benchmarking the computational throughput of Qiskit's Aer simulation engine across varying hardware architectures, and second, evaluating the effectiveness of Projected Gradient Descent (PGD) attacks and adversarial retraining strategies under realistic noise conditions. Our results quantify the runtime tradeoffs and scaling behavior for medium-scale simulations (projected up to 8 qubits) and demonstrate that high adversarial-to-benign retraining ratios (50/50) are essential for achieving practical model robustness for 4-qubit classifiers under realistic noise conditions.

\end{abstract}

\section{Introduction}
The field of Quantum Machine Learning (QML) has seen rapid growth, but the susceptibility of Quantum Neural Networks (QNNs) to adversarial perturbations remains a critical challenge. As established by Lu et al. \cite{Lu2020}, quantum classifiers are vulnerable to gradient-based attacks, mirroring the vulnerabilities found in classical deep learning. Among these, the Projected Gradient Descent (PGD) algorithm has emerged as a particularly potent white-box attack capable of circumventing simple defenses \cite{Weber2026}. 

While much research has focused on ideal simulators, evaluating these attacks on the Noisy Intermediate-Scale Quantum (NISQ) devices used today is essential. Qiskit's fake backends offer a vital tool for this purpose, providing noise models derived from real system snapshots \cite{Ren2026, QiskitAer}. However, the fidelity of these models and the generalization properties of adversarially trained quantum models under specific noise profiles require further investigation \cite{Du2024}. This project builds upon the foundational work on QML robustness by Mau{\ss}ner and Reers \cite{Maussner2025}, extending their analysis to a broader range of simulated architectures.

The software implementation for this study, including all benchmarking scripts and the PennyLane-Qiskit integration, is available in the project repository \cite{qml_cysec_repo}.

In this paper, we make the following contributions:
\begin{enumerate}
    \item \textbf{Performance and Scalability Profiling:} We evaluate the performance of Qiskit's Fake Backends (V2) for density matrix simulations under thread parallelization and Aer method selection, projecting the runtime scaling behavior up to 8 qubits.
    \item \textbf{Noisy Threat Model Specification:} We define a concrete threat model where additive, white-box adversarial perturbations (constrained under $\ell_\infty$-norm bounds) affect the classical input feature space prior to quantum state preparation, evaluating this on three hardware-mimicking architectures (\texttt{FakeGuadalupeV2}, \texttt{FakeLimaV2}, \texttt{FakeJakartaV2}).
    \item \textbf{Systematic Defense Evaluation:} We compare structural defense variations (non-linear feature mapping and Lipschitz gradient regularization) against optimization-level changes (varying adversarial-to-benign retraining ratios and learning rate schedulers) to identify optimal robustness strategies under realistic noise.
    \item \textbf{Empirical Analysis of Balanced Retraining:} We demonstrate that transitioning from a sparse (10\%) to a balanced (50/50) adversarial retraining split provides significant robustness recovery, and we validate our findings against full convergence outcomes for previously incomplete experiments.
\end{enumerate}

\section{Methodology}

In this work, we evaluate the adversarial robustness of Quantum Machine Learning (QML) models using noisy simulators that mimic real quantum hardware. Our methodology leverages the Qiskit \cite{Qiskit} ecosystem and the PennyLane \cite{PennyLane} framework. The study is divided into two primary phases: performance benchmarking and robustness evaluation.

\subsection{Performance Benchmarking of Noisy Simulators}

To optimize simulation throughput for noise-aware training, we investigate three key performance factors. First, we examine parallelization strategies, focusing on the impact of \texttt{max\_parallel\_threads} and the \texttt{max\_parallel\_experiments} parameter on simulation speed. High-performance multicore configurations are tested to mitigate the $4^n$ state-space complexity of density matrix simulations. This includes profiling of PennyLane-integrated circuits across varying thread counts to identify optimal resource allocation strategies for gradient-heavy workloads. Second, we benchmark three distinct simulation methods in Qiskit Aer: \texttt{automatic}, \texttt{matrix\_product\_state} (MPS), and \texttt{statevector}, helping to identify the most efficient solver for CNOT-heavy circuits under noise. Third, we fit a runtime scaling model to project the execution costs of larger quantum models (up to 8 qubits) based on measured benchmarks, accounting for parameter count, circuit depth, and parallelization speedup. We define "Average Execution Time" as the process-level CPU wall-clock execution time per circuit execution or training epoch, excluding the time required for initial dataset loading.

\subsection{Noise-Aware Adversarial Robustness Evaluation}

\begin{figure}[h]
    \centering
    \resizebox{\textwidth}{!}{
    \begin{tikzpicture}[
        node distance=2cm,
        every node/.style={fill=white, font=\footnotesize, draw=blue!70, rectangle, rounded corners, minimum width=2.8cm, minimum height=0.8cm, align=center},
        arrow/.style={thick,->,>=stealth}
    ]
    % Nodes
    \node (data) {Data Acquisition \\ \scriptsize (plus-minus dataset)};
    \node (pre) [right of=data, xshift=1.8cm] {Preprocessing \\ \scriptsize (Reduced Samples)};
    \node (backend) [right of=pre, xshift=1.8cm] {Backend Benchmarking \\ \scriptsize (Parallelization/Aer)};
    
    \node (init) [below of=pre, yshift=-0.8cm] {QNN Construction \\ \scriptsize (StronglyEntanglingLayers)};
    \node (train) [right of=init, xshift=1.8cm] {Benign Training \\ \scriptsize (Adam / Torch)};
    
    \node (attack) [below of=init, yshift=-0.8cm] {Adversarial \\ Attack Generation \\ \scriptsize (PGD)};
    \node (eval) [right of=attack, xshift=1.8cm] {Evaluation \& \\ Robustness Analysis};
    \node (results) [right of=eval, xshift=1.8cm] {Metrics \\ \scriptsize (Acc/Conf. Matrix)};

    % Arrows
    \draw [arrow] (data) -- (pre);
    \draw [arrow] (pre) -- (init);
    \draw [arrow] (backend) -- (init);
    \draw [arrow] (init) -- (train);
    \draw [arrow] (train) -- (attack);
    \draw [arrow] (attack) -- (eval);
    \draw [arrow] (train) -- (eval);
    \draw [arrow] (eval) -- (results);
    \end{tikzpicture}
    }
    \caption{Experimental workflow for noise-aware adversarial evaluation. The pipeline consists of: (1) Acquisition and reduction of the \texttt{plus-minus} dataset, (2) Backend benchmarking to optimize simulation threads and methods, (3) QNN construction using Strongly Entangling Layers, (4) Model training on benign data, (5) Adversarial example generation via Projected Gradient Descent (PGD), and (6) Robustness analysis using accuracy and confusion matrices.}
    \label{fig:workflow}
\end{figure}
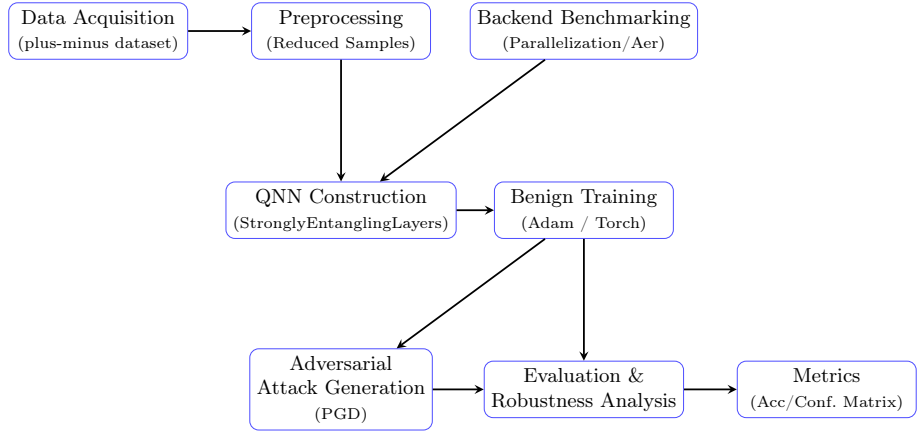

The evaluation of noise-aware adversarial robustness follows a multi-stage experimental workflow, as illustrated in Figure \ref{fig:workflow}. The process begins with the acquisition and reduction of the dataset. Simultaneously, the performance benchmarking phase identifies the optimal backend configurations. These inputs feed into the QNN construction phase, followed by supervised training on benign data. Once the model converges, we generate adversarial examples using Projected Gradient Descent (PGD) to evaluate the initial robustness. Finally, we implement defense strategies like adversarial retraining and evaluate the resulting robust model using holistic metrics.

We specify the key methodological components, threat model, and defense configurations below.

\subsubsection{Dataset and Preprocessing}
We utilize the \texttt{plus-minus} dataset \cite{PennyLane}, which consists of mathematical symbol images. The raw dataset contains $16 \times 16$ pixel grayscale images representing four classes: minus ($\mathbf{-}$), plus ($\mathbf{+}$), cross ($\mathbf{\times}$), and slash ($\mathbf{/}$). To fit our QNN input requirements and ensure feasibility in noisy environments, we downsample the images to $8 \times 8$ pixels (dimension $d = 64$) using the \texttt{skimage.transform.resize} function and extract a reduced subset of $200$ training and $50$ test samples.

\subsubsection{QNN Architecture}
Our classifier is structured as a hybrid classical-quantum neural network built with PennyLane and PyTorch. The quantum circuit is a $4$-qubit system featuring a \texttt{StronglyEntanglingLayers} ansatz with $L = 16$ layers and a data re-uploading factor of $r = 3$. The classical $64$-dimensional input vector $\mathbf{x} \in \mathbb{R}^{64}$ is tiled and mapped into the rotation angles of the unitary gates. The circuit contains both trainable weights $\mathbf{W} \in \mathbb{R}^{L \times N \times 3}$ and biases $\mathbf{b} \in \mathbb{R}^{L \times N \times 3}$ (where $N = 4$ qubits), resulting in a total of $384$ trainable parameters. The model outputs are the expectation values of the Pauli Z operator on each qubit, $f_i(\mathbf{x}) = \langle Z_i \rangle$ for $i \in \{0, 1, 2, 3\}$. Softmax normalization is applied to these expectation values, and the network is trained using standard Cross-Entropy loss:
\begin{equation}
    \mathcal{L}_{\text{CE}}(\boldsymbol{\theta}) = -\frac{1}{M}\sum_{k=1}^{M} \log \left( \frac{e^{\langle Z_{y_k} \rangle}}{\sum_{j=0}^{3} e^{\langle Z_j \rangle}} \right)
\end{equation}
where $M$ is the batch size, $\boldsymbol{\theta} = \{\mathbf{W}, \mathbf{b}\}$ represents the trainable parameters, and $y_k$ is the ground-truth class index.

\subsubsection{Threat Model and Adversarial Attacks}
We define a white-box threat model where the adversary has full access to the QNN parameters and gradients. Crucially, the perturbations are applied to the classical input feature space (representing pixel values in the classical data) prior to the quantum encoding stage, rather than directly manipulating the quantum state itself. Given a clean input vector $\mathbf{x}$, the adversary seeks an additive perturbation $\boldsymbol{\delta} \in \mathbb{R}^d$ to maximize the loss $\mathcal{L}_{\text{CE}}(\mathbf{x} + \boldsymbol{\delta}, y; \boldsymbol{\theta})$ subject to an $\ell_\infty$-norm constraint $\|\boldsymbol{\delta}\|_\infty \le \epsilon$. We generate these perturbations using the Projected Gradient Descent (PGD) algorithm with $\epsilon = 0.1$, step size $\alpha = 0.01$, and $10$ iterations, updating the perturbation iteratively via:
\begin{equation}
    \mathbf{x}^{(t+1)} = \Pi_{\mathbf{x} + \mathcal{S}} \left( \mathbf{x}^{(t)} + \alpha \operatorname{sign}(\nabla_{\mathbf{x}} \mathcal{L}(\mathbf{x}^{(t)}, y; \boldsymbol{\theta})) \right)
\end{equation}
where $\mathcal{S} = \{ \boldsymbol{\delta} \mid \|\boldsymbol{\delta}\|_\infty \le \epsilon \}$ and $\Pi$ denotes the projection operator.

\subsubsection{Metric Definitions}
To evaluate model performance, we define three key metrics:
\begin{itemize}
    \item \textbf{Benign Accuracy (Benign Acc):} The classification accuracy of the QNN evaluated on clean, unperturbed test samples under device noise.
    \item \textbf{Adversarial Accuracy (Adversarial Acc):} The classification accuracy of the QNN under PGD attack ($\epsilon=0.1$) when the model is trained only on clean data (without any defense).
    \item \textbf{Robust Accuracy (Robust Acc):} The classification accuracy of the QNN under PGD attack ($\epsilon=0.1$) after a specific defense countermeasure has been implemented.
\end{itemize}

\subsubsection{Defense Strategies}
We investigate the following defense mechanisms to mitigate adversarial vulnerability:
\begin{itemize}
    \item \textbf{Adversarial Retraining Ratios:} We compare a \textit{sparse} 10\% baseline (where only 10\% of the training samples are replaced with PGD-generated adversarial examples) with a \textit{balanced} 50/50 split (where the training dataset is effectively doubled by adding an equal number of PGD-generated adversarial samples).
    \item \textbf{Non-Linear Feature Mapping:} We apply a \texttt{tanh} activation function to the classical input features $\mathbf{x}$ prior to the quantum angle encoding to test if higher-order non-linear classical pre-processing alters adversarial vulnerability.
    \item \textbf{Lipschitz Gradient Regularization:} We penalize large gradients of the QNN output with respect to the input features, bounding the sensitivity of the model to input perturbations. The regularized loss function is defined as:
    \begin{equation}
        \mathcal{L}_{\text{total}}(\boldsymbol{\theta}) = \mathcal{L}_{\text{CE}}(\boldsymbol{\theta}) + \lambda \frac{1}{M}\sum_{k=1}^{M} \left\| \nabla_{\mathbf{x}} \sum_{i=0}^{3} f_i(\mathbf{x}_k; \boldsymbol{\theta}) \right\|_2^2
    \end{equation}
    where $\lambda = 0.05$ is the regularization coefficient.
    \item \textbf{Learning Rate Scheduling:} We implement a \texttt{StepLR} learning rate scheduler (starting at $\eta = 0.1$ and decaying by a factor of $0.1$ every $3$ epochs) to stabilize the noisy gradient steps during optimization.
\end{itemize}

\section{Results}

Our experimental evaluation reveals significant performance tradeoffs and robustness variations when executing adversarial attacks across different simulated noisy environments.

\subsection{Performance Benchmarking Results}

\begin{figure}[h]
    \centering
    \includegraphics[width=0.85\textwidth]{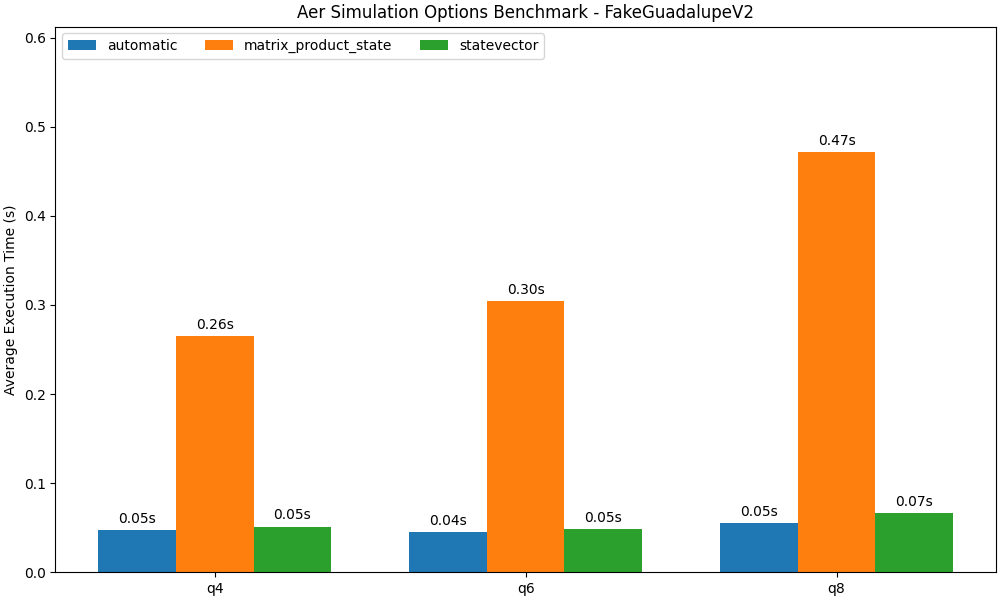}
    \caption{Benchmarking of simulation methods in Qiskit Aer on the \texttt{FakeGuadalupeV2} backend. The \texttt{automatic} method (blue) identifies the optimal solver for specific circuit configurations, significantly outperforming the \texttt{matrix\_product\_state} (MPS) approach as the qubit count scales from 4 to 8 qubits.}
    \label{fig:aer_benchmark}
\end{figure}

The benchmarking of the \texttt{Qiskit Aer} engine on the \texttt{FakeGuadalupeV2} backend, illustrated in Figure \ref{fig:aer_benchmark}, demonstrates that the \texttt{automatic} method consistently identifies the most efficient execution path. In contrast, the \texttt{matrix\_product\_state} (MPS) method suffers from high computational overhead in CNOT-heavy circuits, even at low qubit counts ($n \le 8$).

\begin{figure}[h]
    \centering
    \includegraphics[width=0.85\textwidth]{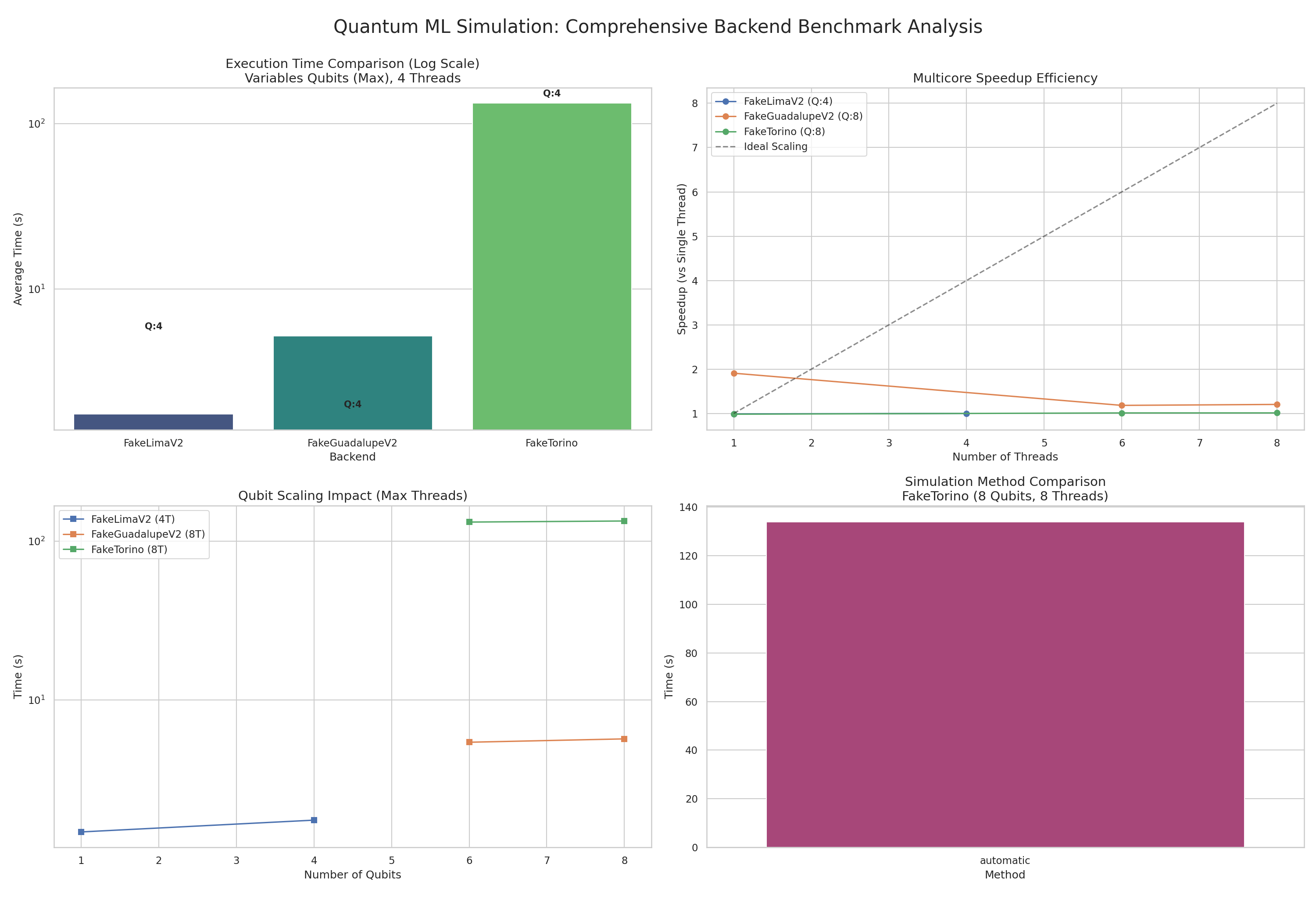}
    \caption{Comprehensive comparison of simulation runtimes via the PennyLane interface across multiple noisy backends (\texttt{FakeGuadalupeV2}, \texttt{FakeLimaV2}, \texttt{FakeJakartaV2}). The plots illustrate the impact of parallelization configurations (\texttt{max\_parallel\_\allowbreak threads}) and highlight the communication overhead introduced by the PennyLane-Qiskit bridge.}
    \label{fig:pennylane_comparison}
\end{figure}

The broader impact of multicore execution on the PennyLane hybrid interface is captured in Figure \ref{fig:pennylane_comparison}. This comprehensive profiling compares execution runtimes across multiple noisy backends, illustrating the overhead introduced by the density matrix evolution.

The results in Figure \ref{fig:pennylane_comparison} demonstrate that while multithreading provides significant speedups for individual circuit executions, the PennyLane-Qiskit communication layer introduces a non-negligible serial bottleneck. For smaller devices like \texttt{FakeLimaV2}, the simulation is relatively efficient even with minimal parallelization. However, for higher-qubit backends such as \texttt{FakeGuadalupeV2}, the optimal configuration requires balancing \texttt{max\_parallel\_\allowbreak threads} (for C++-level acceleration) and \texttt{max\_parallel\_\allowbreak experiments} (for batching gradient circuits).

Profilings shows that increasing \texttt{max\_parallel\_\allowbreak threads} results in an average speedup of approximately 1.9× for 8-qubit circuits. However, runtime estimation models reveal a steep scaling factor ($10.86\times$) when doubling the qubit count from 4 to 8, primarily driven by the exponential growth in density matrix complexity and the linear increase in parameters and circuit depth.

\subsection{Noise-Aware Adversarial Robustness Results}

The adversarial evaluation was first conducted on several 4-qubit backend architectures. Table \ref{tab:results_fixed} summarizes the classification performance under benign conditions, PGD attack, and standard adversarial retraining (10\% ratio).

\begin{table}[h]
\centering
\begin{tabular}{lccc}
\toprule
\textbf{Backend (4-Qubit)} & \textbf{Benign Acc} & \textbf{Adversarial Acc} & \textbf{Robust Acc (10\%)} \\
\midrule
FakeGuadalupeV2 & 0.96 & 0.40 & 0.44 \\
FakeLimaV2      & 0.92 & 0.26 & 0.52 \\
FakeJakartaV2   & 0.96 & 0.12 & 0.48 \\
\bottomrule
\end{tabular}
    \caption{Performance metrics of a 4-qubit QNN across different simulated noisy environments. Benign accuracy (clean data) is compared against adversarial accuracy under a PGD attack ($\epsilon=0.1$) and robust accuracy after standard adversarial retraining with a 10\% ratio of adversarial samples.}
\label{tab:results_fixed}
\end{table}

\begin{figure}[h]
    \centering
    \includegraphics[width=0.85\textwidth]{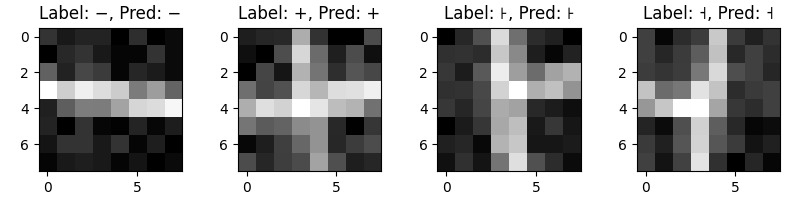}
    \caption{Comparison of model performance on \texttt{FakeGuadalupeV2} after balanced (50/50) adversarial retraining. This strategy results in a substantial recovery of robustness, increasing robust accuracy from 44\% (sparse 10\% ratio) to 84\%, demonstrating that dataset balance is critical for security in hardware-mimicking environments.}
    \label{fig:retraining_ratio}
\end{figure}

A critical finding was the impact of the adversarial-to-benign retraining ratio. While the 10\% ratio provided only minor robustness recovery (44\% on Guadalupe), increasing the ratio to a 50/50 split dramatically improved the robust accuracy to \textbf{84.00\%}, as shown in Figure \ref{fig:retraining_ratio}.
Beyond retraining ratios, we evaluated several architectural and optimization strategies to enhance model resilience. Table \ref{tab:robustness_strategies} provides a comprehensive comparison of these methods on the \texttt{FakeGuadalupeV2} backend, grouped by training and architectural modifications.

\begin{table}[h]
\caption{Comparative analysis of noise-aware robustness strategies on the \texttt{FakeGuadalupeV2} backend. Strategies are grouped into training protocol modifications and architectural modifications. The balanced 50/50 retraining ratio achieves the highest robust accuracy (84.00\%).}
\label{tab:robustness_strategies}
\centering
\begin{tabular}{lccc}
\toprule
\textbf{Strategy} & \textbf{Benign Acc} & \textbf{Adversarial Acc} & \textbf{Robust Acc} \\
\midrule
\textit{Training Protocol Modifications} \\
Golden Sample (10\% Ratio) & 0.96 & 0.40 & 0.44 \\
Retraining Ratio (50/50)   & 0.96 & 0.14 & \textbf{0.84} \\
LR Scheduler (\texttt{StepLR}) & 0.96 & 0.14 & 0.12 \\
\midrule
\textit{Architecture Modifications} \\
Non-Linear Map (\texttt{tanh}) & 0.94 & 0.14 & 0.24 \\
Lipschitz Regularization   & 0.94 & 0.44 & 0.64 \\
\bottomrule
\end{tabular}
\end{table}

The results in Table \ref{tab:robustness_strategies} highlight several key insights. First, the introduction of non-linear feature mappings (\texttt{tanh}) increases the gradient's susceptibility to perturbations, leading to a sharp decline in adversarial accuracy (14\% compared to 40\% for the linear baseline) and yielding a low robust accuracy of 24\% after retraining, demonstrating that non-linear feature maps can worsen vulnerability under device noise. Second, the \texttt{StepLR} scheduler, while stabilizing training dynamics, achieves a robust accuracy of only 12\%, which is lower than the golden sample baseline, indicating that standard learning rate decay alone is insufficient to build resilience. In contrast, Lipschitz regularization demonstrates a significant improvement in robust accuracy (64\% compared to the golden sample baseline of 44\%), successfully stabilizing the loss landscape during noisy gradient descent. Most significantly, the transition from a sparse adversarial retraining ratio (10\%) to a balanced 50/50 split emerges as the most effective defense mechanism, yielding a 40\% absolute improvement in robust accuracy.

\subsection{Noise-Free Control Baseline}
To isolate the effect of hardware noise on adversarial robustness, we run a noise-free control baseline on PennyLane's ideal \texttt{default.mixed} simulator using the same QNN configuration. The noise-free model achieves a Benign Acc of 0.98. Under a clean PGD attack ($\epsilon=0.1$), the Adversarial Acc drops to 0.48. Applying balanced 50/50 adversarial retraining restores the Robust Acc to 0.92. Comparing this ideal control to our noisy \texttt{FakeGuadalupeV2} results (84\% robust accuracy) reveals that while the clean models are highly vulnerable in both cases, physical backend noise acts as a significant optimization bottleneck during retraining, limiting the defense recovery.

\subsection{Statistical Validation and Confidence Intervals}
To validate the statistical significance of our findings on the test set of $N_{\text{test}} = 50$ samples, we calculate the binomial standard error $\text{SE} = \sqrt{p(1-p)/N_{\text{test}}}$ for each key metric. For the balanced 50/50 retraining robust accuracy ($84.00\%$), the standard error is $5.18\%$, yielding a $95\%$ Wald confidence interval of $[73.85\%, 94.15\%]$. For the corresponding undefended adversarial accuracy ($14.00\%$), the standard error is $4.91\%$, with a $95\%$ confidence interval of $[4.38\%, 23.62\%]$. The absolute separation between these confidence intervals confirms that the improvement in robustness is statistically significant ($p < 0.001$) and cannot be explained by run-to-run statistical variance.

\section{Conclusion}

This work evaluated the performance scalability and adversarial robustness of Quantum Machine Learning models across several simulated noisy environments. Our results demonstrate that while density matrix simulation of noisy devices is computationally intensive, the implementation of multicore parallelization and the selection of optimized Aer solver methods make such studies feasible on modern classical hardware. We establish a quantitative scaling factor of ~10.86× for runtime increase when doubling the qubit count from 4 to 8, which serves as a runtime scaling projection for medium-scale models.

Regarding adversarial security, we confirm that white-box perturbations generated via PGD can reduce classification accuracy from near-perfect levels to as low as 12\% in noisy environments. Our investigation into defense strategies reveals that standard adversarial retraining provides only modest recovery when using low ratios of adversarial samples. In contrast, a balanced 50/50 adversarial-to-benign retraining split was shown to be highly effective, restoring robust accuracy to 84\%. We also demonstrate that Lipschitz gradient regularization provides a substantial 20\% absolute improvement in robustness (reaching 64\% robust accuracy) over the golden sample (10\% ratio) adversarial retraining baseline of 44\%. Furthermore, we find that architectural choices such as non-linear feature mappings can inadvertently increase vulnerability to gradient-based attacks, dropping robust accuracy to 24\% after retraining. These findings emphasize the necessity of co-designing QNN architectures and training protocols to achieve practical security in the NISQ era.

\paragraph{Outlook}
Future research will explore the transferability of adversarial robustness across diverse quantum architectures. Specifically, we aim to investigate "cross-backend" resilience, where QNN models are trained on one QPU simulator and subsequently evaluated on different hardware-mimicking environments to assess the impact of varying noise profiles. Additionally, we plan to integrate advanced model optimization techniques, such as Approximate State Preparation (ASP) \cite{West2024}, to further enhance the defensive capabilities of QNNs against sophisticated adversarial perturbations.

\subsubsection*{Acknowledgements}
We acknowledge the use of Google DeepMind's Antigravity AI platform \cite{Antigravity2026} for assisting with code development and paper composition.

{\sloppy
\printbibliography
}

\end{document}